\documentclass{ptephy}

\usepackage{graphicx}
\usepackage{amssymb}
\usepackage{color} 
\begin{document}

\title{Search for inelastic WIMP nucleus scattering
on $^{129}$Xe in data from the XMASS-I experiment}

\author{\name{\fname{H.~Uchida}}{8,\ast},
\name{\fname{K.~Abe}}{1,3},
\name{\fname{K.~Hieda}}{1},
\name{\fname{K.~Hiraide}}{1,3},
\name{\fname{S.~Hirano}}{1},
\name{\fname{K.~Ichimura}}{1},
\name{\fname{Y.~Kishimoto}}{1,3},
\name{\fname{K.~Kobayashi}}{1,3},
\name{\fname{S.~Moriyama}}{1,3},
\name{\fname{K.~Nakagawa}}{1},
\name{\fname{M.~Nakahata}}{1,3},
\name{\fname{H.~Ogawa}}{1,3},
\name{\fname{N.~Oka}}{1},
\name{\fname{H.~Sekiya}}{1,3},
\name{\fname{A.~Shinozaki}}{1},
\name{\fname{Y.~Suzuki}}{1,3},
\name{\fname{A.~Takeda}}{1,3},
\name{\fname{O.~Takachio}}{1},
\name{\fname{D.~Umemoto}}{1},
\name{\fname{M.~Yamashita}}{1,3},
\name{\fname{B.~S.~Yang}}{1},
\name{\fname{S.~Tasaka}}{2},
\name{\fname{J.~Liu}}{3},
\name{\fname{K.~Martens}}{3},
\name{\fname{K.~Hosokawa}}{5},
\name{\fname{K.~Miuchi}}{5},
\name{\fname{A.~Murata}}{5},
\name{\fname{Y.~Onishi}}{5},
\name{\fname{Y.~Otsuka}}{5},
\name{\fname{Y.~Takeuchi}}{5,3},
\name{\fname{Y.~H.~Kim}}{6},
\name{\fname{K.~B.~Lee}}{6},
\name{\fname{M.~K.~Lee}}{6},
\name{\fname{J.~S.~Lee}}{6},
\name{\fname{Y.~Fukuda}}{7},
\name{\fname{Y.~Itow}}{8,4},
\name{\fname{K.~Masuda}}{8},
\name{\fname{Y.~Nishitani}}{8},
\name{\fname{H.~Takiya}}{8},
\name{\fname{N.~Y.~Kim}}{9},
\name{\fname{Y.~D.~Kim}}{9},
\name{\fname{F.~Kusaba}}{10},
\name{\fname{K.~Nishijima}}{10},
\name{\fname{K.~Fujii}}{11},
\name{\fname{I.~Murayama}}{11},
\name{\fname{S.~Nakamura}}{11}}
%

\address{
\affil{1}{Kamioka Observatory, Institute for Cosmic Ray Research, the University of Tokyo, Higashi-Mozumi, Kamioka, Hida, Gifu, 506-1205, Japan}
\affil{2}{Information and multimedia center, Gifu University, Gifu 501-1193, Japan}
\affil{3}{Kavli Institute for the Physics and Mathematics of the Universe (WPI), the University of Tokyo, Kashiwa, Chiba, 277-8582, Japan}
\affil{4}{Kobayashi-Maskawa Institute for the Origin of Particles and the Universe, Nagoya University, Furo-cho, Chikusa-ku, Nagoya, Aichi, 464-8602, Japan}
\affil{5}{Department of Physics, Kobe University, Kobe, Hyogo 657-8501, Japan}
\affil{6}{Korea Research Institute of Standards and Science, Daejeon 305-340, South Korea}
\affil{7}{Department of Physics, Miyagi University of Education, Sendai, Miyagi 980-0845, Japan}
\affil{8}{Solar-Terrestrial Environment Laboratory, Nagoya University, Nagoya, Aichi 464-8601, Japan}
\affil{9}{Department of Physics, Sejong University, Seoul 143-747, South Korea}
\affil{10}{Department of Physics, Tokai University, Hiratsuka, Kanagawa 259-1292, Japan}
\affil{11}{Department of Physics, Faculty of Engineering, Yokohama National University, Yokohama, Kanagawa 240-8501, Japan}
\email{xmass.publications@km.icrr.u-tokyo.ac.jp}}

\begin{abstract}%
A search for inelastic scattering of Weakly Interacting Massive Particles
(WIMPs) on the isotope $^{129}$Xe was done
in
data taken with the single
phase liquid xenon detector XMASS at the Kamioka Observatory. Using
a restricted volume containing 41\,kg of LXe at the very center of our detector
we observed no significant
excess of events
in 165.9\,live\,days of data.
Our background reduction allowed us to derive our limits without
explicitly subtracting the remaining events which are compatible
with background expectations.
We derive for e.g.\ a 50\,GeV WIMP an upper
limit for its inelastic cross section on $^{129}$Xe nuclei of 3.2\,pb at the
90\% confidence level. 
\end{abstract}

\subjectindex{C43, E64, F41}

\parindent0pt

\maketitle

\section{Introduction}
There is ample observational evidence for the existence of dark matter in
the Universe. All evidence is gravitational though: Observing the distribution
and motion of normal matter in galaxies and clusters we derive its presence.
No evidence exists for any interaction other than the gravitational one
that dark matter might have with normal matter.
Theory provides strong motivation to postulate WIMP dark matter
though.
If WIMPs do
indeed make up the bulk of the dark matter in the universe and its weak
coupling to normal matter is strong enough, elastic scattering of nuclei
should provide the experimental signature for such an interaction.
Unfortunately the resulting recoil energy is very low ($\sim$10\,keV), and the
spectrum of the recoiling nuclei is falling off exponentially, making it
difficult to distinguish from backgrounds near
detector
threshold.

All around the world
significant experimental effort is expended to probe for such 
nuclear recoils through elastic scattering
\cite{dama,dama_2,cogent,cresst,xenon,xenon_2,cdms,cdms_2,edelweiss}.
Inelastic scattering that excites
low lying nuclear states in suitable target nuclei
provides another avenue to probe for WIMP dark matter. Its
advantage is that nuclear excited states and their de-excitation mechanisms
are typically well measured, and thus the expected energy deposit in the
detector is known, resulting in the readily identifiable signature of a line
in the energy spectrum.

Experimental searches were done with $^{127}$I,
which has a suitable nuclear excitation level at 57.6\,keV
\cite{Ejiri, Fushimi}.
Among the xenon isotopes found in naturally occurring xenon
$^{129}$Xe has the lowest lying excited nuclear state at
39.58\,keV and with 26.4\% has almost the highest natural
abundance; the runner up would be $^{131}$Xe with 21.2\%
abundance and an 80.19\,keV excitation. Thus the 
$^{129}$Xe excitation threshold is
lower than that for $^{127}$I,
yet significantly above both the XMASS data acquisition and
analysis thresholds.
The de-excitation of this M1 state in $^{129}$Xe proceeds through gamma ray emission or an internal conversion electron with subsequent X-ray emission. 
With its high nuclear charge Xe itself is a good absorber for
such gamma rays, providing liquid xenon (LXe)
detectors with an intrinsically high
detection efficiency for the prospective signal.

So far the DAMA group searched for this signal
in a 2500\,kg$\cdot$day exposure of
6.5\,kg of LXe. They used 99.5\% enriched $^{129}$Xe and constrained the inelastic
cross section for 50\,GeV WIMPs to be less than 3\,pb at 90\% confidence level
(C.L.) \cite{dama1996,dama2000}.

In this paper results from our own search for this signal in XMASS data is
reported. Though the LXe in XMASS contains $^{129}$Xe only at the level
of its natural abundance, our detector's significantly lower background 
in its fiducial volume, which is a spherical volume around
the center of the detector,
and
excellent light yield
result in a high sensitivity for
this
inelastic scattering signal.

\section{The XMASS detector}
The XMASS experiment is located underground in the Kamioka Observatory at a depth of 2700\,m.w.e., aiming to detect dark matter \cite{suzuki}.  
XMASS is a single phase liquid xenon scintillation detector containing 1050\,kg of Xe in its OFHC copper vessel.
Xenon scintillation light is detected by 642 inward-pointing Hamamatsu R10789 series photomultiplier tubes (PMTs) arranged on an 80\,cm diameter pentakis-dodecahedron support structure within the LXe containment vessel to give a total photocathode coverage of 62.4\,\% of the detector\rq{}s inner surface. 
This surface encloses an active target region containing 835\,kg of liquid xenon.

To shield the scintillator volume from external gamma rays and neutrons, and to veto muon-induced backgrounds, this active 
target of our detector
is located at the center of a $\phi$\,10\,m $\times$ 11\,m cylindrical tank filled with pure water.
This 
water
volume is viewed by 72 Hamamatsu R3600 20-inch PMTs to provide an active muon veto as well as being a passive shield against external backgrounds.
This is the first
time a
water Cherenkov shield
is
used in
a direct dark matter search.

Radioactive calibration sources can be inserted through
a portal above the center of the detector and be positioned
along the central vertical axis of the inner detector to
calibrate energy as well as position reconstruction.
Measuring with a $^{57}$Co source from the center of the detector volume the photoelectron yield\footnote{This photoelectron yield is smaller than the value  reported
 in Ref.~\cite{xmass_det, xmass_sci, xmass_sci_2} since we changed a correction on the
 charge observed in our electronics. This correction is within
 the uncertainty reported earlier~\cite{xmass_det}.} is determined to be 13.9 photoelectrons (p.e.)/keV.
A more detailed description of the XMASS detector is
presented in~\cite{xmass_det}.

PMT signals are passed though preamplifiers with
a gain of 11
before being processed by Analog-Timing-Modules (ATMs)~\cite{sk}.  
These modules combine the functions of typical ADC and TDC modules, recording both the integrated charge and the arrival time of each PMT signal.  
For each PMT channel the discriminator threshold is set to $-5$\,mV, which corresponds to $0.2$ p.e.. 
When a PMT signal exceeds this threshold, a \lq\lq{}hit\rq\rq{}
is registered on the
channel.
A global trigger is generated if the number of hit PMTs
within a 200\,ns window is more than nine.

A complete XMASS detector Monte Carlo (MC) simulation package based on Geant4~\cite{geant4,geant4_2} including
a simulation
of 
the readout electronics has been developed~\cite{xmass_det}
and is used in our analysis.
The simulation has been tuned using calibration data and the optical 
properties of the liquid xenon have also been extracted from
calibration data.
The energy dependence of the light yield as well as the
energy resolution were also tuned on calibration data,
as were the decay constant of gamma induced scintillation light
and the transit timing spread (TTS) of the PMTs.
We choose these constants
so that we can reproduce the observed distribution of PMT hit timings
in our simulation. The effective decay constant $\tau_\gamma$ thus determined is 27.3\,ns for a 39.58\,keV gamma ray and the TTS is 2.33\,ns (rms).
The manufacturer evaluated
TTS for our PMTs is 2.4\,ns.
For $\tau_\gamma$ Ref.~\cite{Akimov} reports 34\,ns but
as our simulation reproduces the observed timing distribution
for various gamma sources at both lower and higher energies,
we do not consider this only known outside measurement
in our evaluation of systematic uncertainty.

\section{Expected Signal and Detector Simulation}
WIMP on $^{129}$Xe inelastic scattering produces
a 39.58\,keV $\gamma$-ray from
nuclear
de-excitation
plus a few keV energy deposition
from the recoil of the 
$^{129}$Xe nucleus.
Energy spectra for the nuclear recoil part are obtained
by simulation, just
as in Ref.\ \cite{dama1996,dama2000}. The differential rate
for inelastic scattering of WIMPs on nucleons here
as in these references is calculated according to:
\begin{equation}
  \frac{dR}{dE_{det}} =\frac{dE_{nr}}{d(\mathcal{L}_{\rm eff}E_{nr})}\frac{dR}{dE_{nr}} =\frac{dE_{nr}}{d(\mathcal{L}_{\rm eff}E_{nr})}\frac{\rho_WN_T\sigma_{I}^{as}M_Nc^2}{2M_W\mu^{2}} F^{2}(E_{nr}) \int_{v_{min}(E_{nr})}^{v_{max}} \frac{1}{v}\frac{dn}{dv}dv,
\end{equation}
where $R$ is event rate in a unit mass of the target, $E_{det}$
is the detected energy in electron equivalent
deposited energy,
$E_{nr}$ is the
nuclear
recoil energy, 
$\mathcal{L}_{\rm eff}=\mathcal{L}_{\rm eff}(E_{nr})$
is a factor that converts nuclear recoil energy $E_{nr}$ to electron
equivalent energy $E_{det}$ relative to that of 122\,keV gamma at
zero electric field \cite{Leff, xmass_sci, Leff2, Leff3}, 
$\rho_W$ is the local mass density of dark matter
(0.3\,GeV/cm$^3$) \cite{PDG},
$N_T$ is the number of target nuclei, $\sigma^{as}_I$ is the asymptotic
cross section for inelastic scattering
at zero momentum transfer,
$M_N$ is the mass of the
target nucleus, $M_W$ is the WIMP mass, $\mu$ is the reduced mass
of the WIMP mass and the target nucleus mass,
$F^2(E_{nr})$ is the nuclear form factor of $^{129}$Xe, $v_{max}$ is the
maximum velocity of the WIMPs in the Earth's vicinity
(approximated by the local escape velocity for the galaxy, 650\,km/s),
$v_{min}(E_{nr})$ is the minimum
velocity
the WIMP must have to be able to excite a nucleus,
$v$ is the velocity
of the WIMP, and $dn/dv$ is the velocity distribution of WIMPs.
The velocity distribution, $dn/dv$, is assumed to be quasi-Maxwellian with the most probable thermal speed of the WIMPs
being
$v_0$=220\,km/s \cite{Jungman}, and the average velocity of the Earth in the galactic
frame $v_e$=232\,km/s \cite{Freese}.
Following Ref.\ \cite{dama1996,dama2000}
the minimum velocity needed to excite $^{129}$Xe is
evaluated as:
\begin{eqnarray}
v_{min} = v_{min}^{0} + \frac{v_{thr}^{2}}{4v_{min}^{0}},
\end{eqnarray}
with: 
\begin{eqnarray}
  v_{min}^{0} = \sqrt{\frac{M_N E_{nr}}{2\mu^2}} \\
  E_{det} = E^{\ast} + \mathcal{L}_{\rm eff}E_{nr} \\
  v_{thr}^{2} = 2\Delta Ec^2/\mu,
\end{eqnarray}
where $\Delta E$ is the energy of the first excited state
of $^{129}$Xe (39.58\,keV) and $E^{\ast} \sim \Delta E$
is the sum of all the energy deposited in the de-excitation process.

The total event rate in the case of a point-like target thus becomes:
\begin{equation}
R_{I\rm, point\mathchar`-like} = \int_{v_{thr}}^{v_{max}}\frac{\rho_Wv}{M_W}N_T\sigma_I(v)\frac{dn}{dv}dv = \frac{\rho_W{\langle}v{\rangle}}{M_W}fN_T\sigma_{I}^{as}
\end{equation}
\begin{equation}
f = \frac{1}{{\langle}v{\rangle}}\int_{v_{thr}}^{v_{max}}(v^2-v_{thr}^2)^{1/2}\frac{dn}{dv}dv,
\end{equation}
where $\sigma_I(v)$ is the excitation cross section for
a point-like target, which is 
expressed as the following function of the WIMP velocity $v$:
\begin{equation}
\sigma_I(v) = \frac{\mu^2}{{\pi}M_N}|{\langle}N^*|M|N{\rangle}|^2{\Bigl(}1-\frac{v_{thr}^2}{v^2}{\Bigr)}^{1/2} = \sigma_I^{as}{\Bigl(}1-\frac{v_{thr}^2}{v^2}{\Bigr)}^{1/2},
\end{equation}
with ${\langle}N^*|M|N{\rangle}$ being the matrix element
for inelastic scattering;
details can be found in Ref.\ \cite{Ellis}.
To incorporate effects of the finite size of the $^{129}$Xe nucleus, 
the form factor $F^2(E_{nr})$ should be taken into account. 
In fact, there is significant progress with regard to
form factors since the DAMA results were published
\cite{Engel, Toivanen, Toivanen_2, Klos, Baudis, Ejiri2013}.
In this paper, we first choose the model of Ref.~\cite{Engel},
which is the same as used by the DAMA group \cite{dama1996,dama2000}
to allow for comparison with this only other inelastic scattering result
for $^{129}$Xe.
Next we use the more recent calculations
in Refs.~\cite{Baudis, Ejiri2013} for interpreting
our results in terms of a constraint
on the spin dependent WIMP-neutron cross section,
which can be compared with results
from elastic scattering
determinations of that cross-section.

\begin{figure}[]
\begin{tabular}{cc}
\begin{minipage}{75mm}
    \begin{center}
\includegraphics[trim=0cm 0cm 0cm 0cm,width=7.5cm,height=7.5cm,angle=0,keepaspectratio,clip]{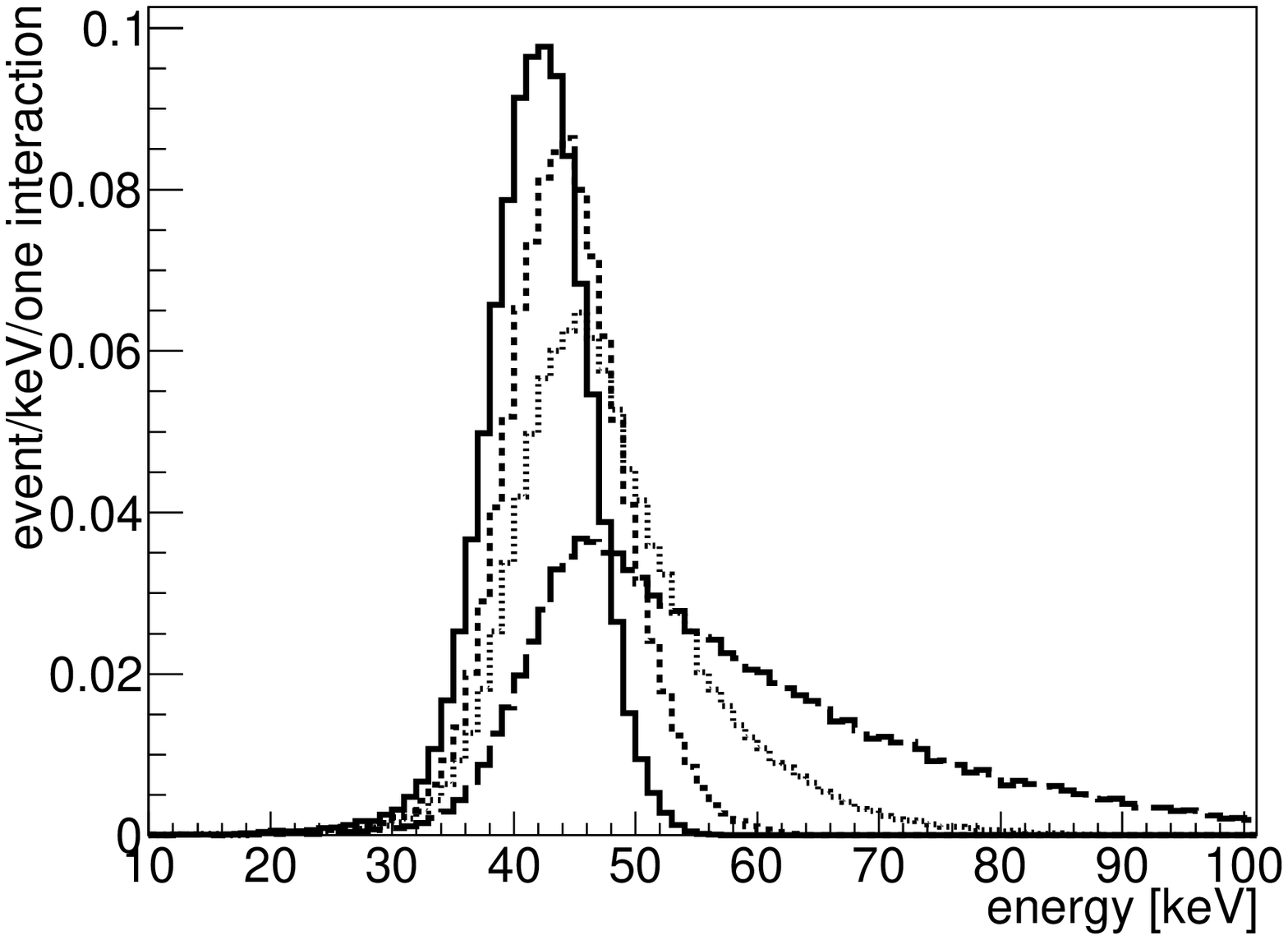}
     \end{center}
\end{minipage} &
\begin{minipage}{75mm}
\begin{center}
\includegraphics[trim=0cm 0cm 0cm 0cm,width=7.5cm,height=7.5cm,angle=0,keepaspectratio,clip]{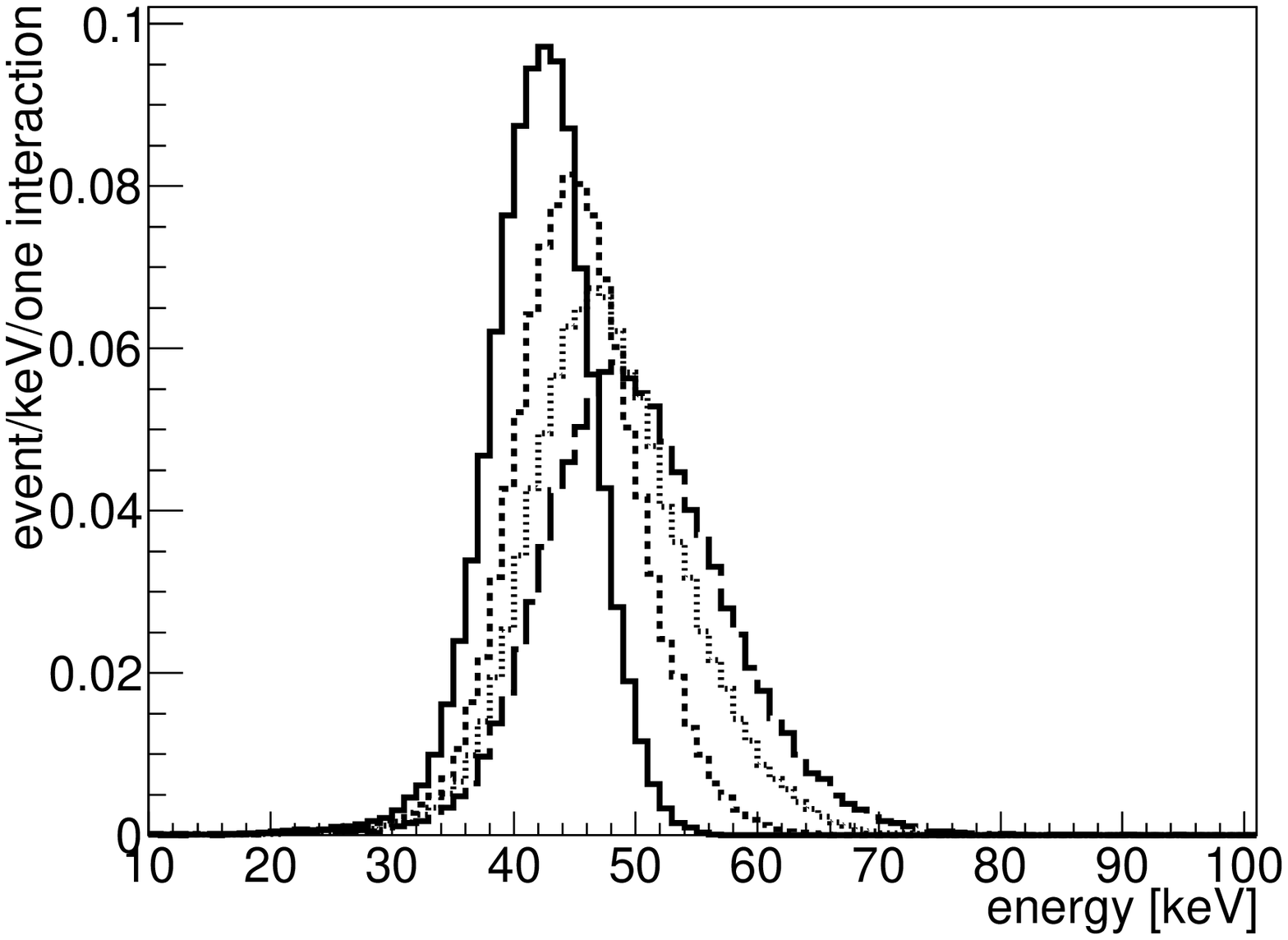}
     \end{center}
\end{minipage} 
\end{tabular}
     \caption{Simulated energy spectra for WIMP
	 masses of
     20 (solid line), 50 (dashed line), 100 (dotted line),
     and 1000\,GeV (long dashed line). 
     (a) Calculated with the form factor in Ref.~\cite{Engel} and
     (b) with the form factor in Ref.~\cite{Baudis}.
       In both cases it includes the energy deposited
       by the de-excitation gamma ray as well as
       the energy deposit effected by the nuclear
       recoil. Our energy scale is defined
     by dividing the number of p.e. observed
     by 13.9 (p.e./keV).}
    \label{fig:1}  
\end{figure}

The expected signals from both de-excitation and the associated nuclear recoil
are simulated and then added.
In the simulation, scintillation light
emission due to nuclear recoil
(decay constant $\tau_{nr}=25$\,ns \cite{Ueshima})
and the subsequent $\gamma$-ray 
or conversion electron emission from nuclear
de-excitation
are simulated at their common vertex in the detector. Here we ignore the few percent difference of K shell and L shell electron ejection probability
after de-excitation and gamma ray absorption.
The vertices are distributed uniformly throughout the inner detector.
The half life of the excited
nuclear
state can be ignored since
it is much shorter ($\sim$1\,ns) than the decay constant
of the scintillation light.
Figure~\ref{fig:1} shows the resulting simulated
energy deposits for WIMPs with various masses (a) with the form factor
from Ref.~\cite{Engel} and (b) with the form factor from Ref.~\cite{Baudis}.

\section{Data reduction and Optimization}
The data used for this search was taken 
between December 24, 2010 and May 10, 2012, and amounts to a total live time of 
165.9 days.
Since we took extensive calibration
data and various special runs to understand the background and the general 
detector response, we select 
runs taken under what we designate \lq\lq{}normal running conditions\rq\rq{} with stable 
temperature and pressure (0.160-0.164\,MPa absolute).
Additional data quality checks reject runs with excessive PMT noise,
unstable pedestal levels, or abnormal trigger rates.

As discussed in Ref.\ \cite{xmass_sci,xmass_sci_2},
most of the observed events are background events
due to radioactive contamination in the aluminum seals of our PMTs
and radon progeny on the inner surface of the detector.
To reduce these backgrounds, 
a dedicated event reduction procedure was developed for this analysis.
In this section we give a detailed description of this reduction procedure and our evaluation of its acceptance.

This dedicated data reduction proceeds in four steps:
(1) Pre-selection. This is similar to
Ref.\ \cite{xmass_sci}.
The difference is that events occurring less than 10\,ms
prior to 
the one under consideration are also
rejected since events caused by $^{214}$Bi decay, a daughter
of $^{222}$Rn, must be removed.
(2) Fiducial volume (radius) cut. As is described in Ref.\ \cite{xmass_det},
the observed pattern of p.e. is used to reconstruct an event vertex.
The radial position $R$ of an event is obtained from this reconstruction.
(3) Timing cut. Even after the radius cut, some surface events
remain in the sample. Timing information
is used to further reduce these remaining surface background events.
Here we use the timing difference $\delta T_m$
between the first hit and
an average of hit timings of first 50\%
of an events' remaining hits-after discarding
the next ten following that first one.
A larger timing difference is indicative of a surface event
that was mistaken for a fiducial volume event.
(4) Band cut. Grooves and gaps exist between PMTs.
Scintillation light caused by events inside those grooves
projects onto the inner surface of the XMASS detector 
in a characteristic band pattern.
This pattern emerges because the propagation of scintillation
light from within a
straight groove is constrained by the rims of that groove
acting as a slit projecting
a characteristic band shape that is recognized by our software.
Events with such a pattern are
eliminated by this cut.
Using the band identified by our software we cut
on the ratio of charges contained in that
band to the total charge in the event:
\begin{equation}
\mbox{Band cut parameter $F_B$} = \frac{\mbox{p.e. in the band of width 15\,cm}}{\mbox{Total p.e. in the event}}
\end{equation}

The cut values for
the three cuts
which are applied after our almost standard pre-selection
were optimized for a WIMP mass of 50\,GeV.
Except for the radius cut our cut values were determined
by optimizing the ratio of simulated signal events
surviving the cuts in a tentative signal range
from 30 to 80\,keV 
over the sum of background events found in the data in two side
bands ranging from 10 to 30 and from 80 to 100\,keV.
For the radius cut this procedure results in an extremely
low fiducial volume, leading us to halt this optimization at 15\,cm.
For the remaining cuts the values resulting from our optimization
were 12.91\,ns for the timing cut and a ratio of 0.248 for the band cut.
Events with parameter values smaller than these cut values
enter into the final sample.

\begin{figure}[]
    \begin{center}
\includegraphics[trim=0cm 0cm 0cm 0cm,width=10cm,height=10cm,angle=0,keepaspectratio,clip]{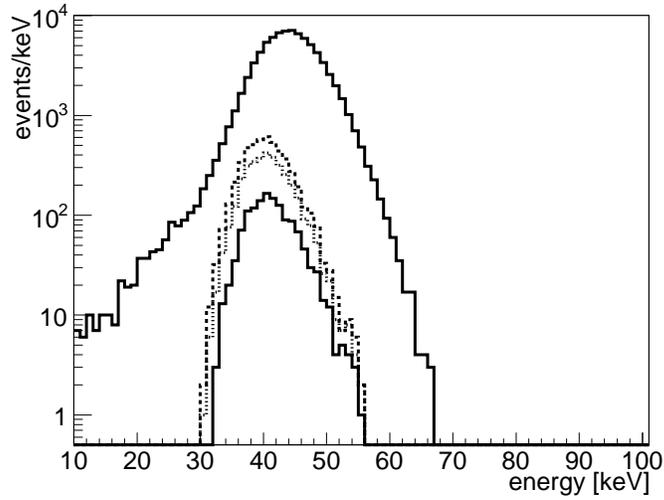}
     \end{center}
     \caption{
     Energy spectra of the simulated events
     after each reduction step.
     As an example we chose a WIMP mass of 50\,GeV
     and the form factor in Ref.~\cite{Engel}.
     From top to bottom, simulated energy spectrum
     after the pre-selection (solid line),
     cut (2) (dashed line), 
     cut (3) (dotted line),
     and cut (4) (solid line).
     As we do not apply the proper radial correction
     for energy, a shift in our energy scale seems to occur
     after our fiducial volume cut (2).
     As we are only using events
     in a very limited fiducial volume and our energy
     scale is based on calibration at the center of the
     detector, the energy scale of the surviving events
     is correct within 4\%.
     }
    \label{fig:2}
\end{figure}

\begin{figure}[]
    \begin{center}
\includegraphics[trim=0cm 0cm 0cm 0cm,width=10cm,height=10cm,angle=0,keepaspectratio,clip]{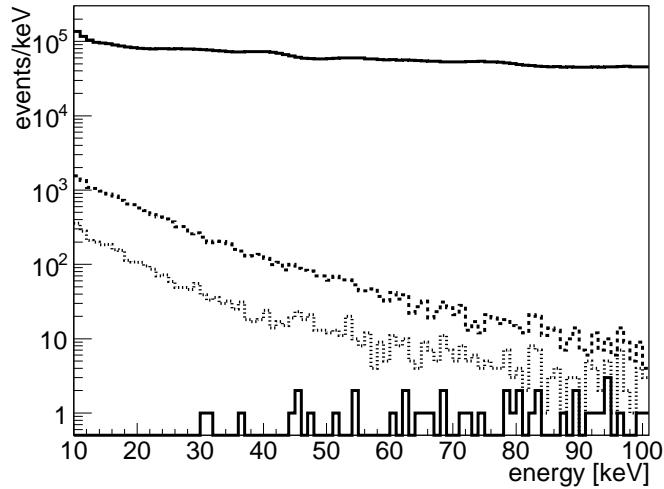}
     \end{center}
     \caption{Energy spectra of the observed events
     after each reduction step for our 
     165.9\,live\,days data.
     From top to bottom, the observed energy spectrum
     after the pre-selection (solid line),
     cut (2) (dashed line), 
     cut (3) (dotted line),
     and cut (4) (solid line).
     The fiducial volume contains 41\,kg of LXe.
     }
    \label{fig:3}
\end{figure}

Figures~\ref{fig:2} and \ref{fig:3} show the impact of our cuts on
the expected signal from our 50\,GeV WIMP simulation and
the observed data spectrum, respectively.
The signal window is defined
so that it contains 90\% of the simulated
50\,GeV WIMP signal with equal 5\% tails to either side,
which results in a 36-48\,keV window.
While the underlying simulation shown in Fig.~\ref{fig:2}
is based on the form factors used in Ref.~\cite{Engel},
it can be seen in Fig.~\ref{fig:1} that the shape of
this distribution for a 50\,GeV WIMP does not change much
with the use of the more modern form factors.
These cut values and the signal window optimized for the 50\,GeV WIMPs were also used to obtain the limits for the other WIMP masses.
Our signal efficiency is defined as the 
ratio between the number of simulated events remaining after all cuts
in the 36-48\,keV signal region and the number of simulated events generated
within the fiducial volume (radius less than 15\,cm,
containing 41\,kg of LXe).
As shown in Tab.~\ref{gel1} signal efficiency ranges from 
29\% for 50\,GeV WIMPs to 15\% for 5\,TeV WIMPs
for the nuclear form factors given in Ref.~\cite{Engel}.

\section{Results and Discussion}
As clearly visible in Fig.~\ref{fig:3},
the cuts discussed in the previous section almost
eliminate all background in and around the signal
window.
After all cuts 5 events are remaining in our 36-48\,keV signal region.
The main contribution to the remaining background in this energy region
stems from the $^{222}$Rn daughter $^{214}$Pb.
From our simulation we estimate this background alone to contribute
2.0$\pm$0.6\,events. As other background contributions are smaller
but less certain, we do not subtract background when calculating
our limits.
Our detector's low background allows us to
directly use the event count in the signal region
to extract our limit on the inelastic scattering cross section
of WIMPs on $^{129}$Xe nuclei.
Using Eq.~6 and taking into account the nuclear form factor and our
signal efficiency we derive the 90\% C.L.\ upper limit for this cross section
which in Fig. 4 is compared to the result from \cite{dama1996,dama2000}. The gray band reflects our systematic uncertainties.
The systematic uncertainty on our signal efficiency is estimated
from data-MC comparisons for $^{241}$Am calibration data
(60\,keV) at various positions within the fiducial volume.
The relevant comparisons are shown in Figs.~\ref{fig:5} and \ref{fig:6}.
From these comparisons we derive the systematic uncertainties in energy
scale, energy resolution, radius reconstruction, timing cut,
and band cut parameter. 
There is uncertainty also in the effective light yield $\mathcal{L}_{\rm eff}$
and the decay constants of nuclear recoils.
The cumulative effect of these individual contributions
is obtained by summation in quadrature.

As an example for our systematic error evaluation we explain it 
here
for the signal efficiency for 50\,GeV WIMPs;
see Tab.~\ref{gel2} for other masses.
The uncertainty in our energy scale evaluates to
$\pm^{4.6}_{3.1}$\% by comparing more than 10 sets of calibration data 
($^{57}$Co), taken at different times throughout
the data taking period, with our simulation.
Changing the number of photons generated per unit energy deposited
in the simulation by this amount, the signal efficiency changes 
by $\pm^{11}_{13}$\%.
The uncertainty in the energy resolution, 12\%, is evaluated
by comparing the resolution of the 60\,keV peak
in the calibration data and simulated events.
This leads to a 5.6\% reduction in the signal efficiency.
The radial position of the reconstructed vertex for the 
calibration data differs by 5\,mm from the true source position,
which causes a 3.2\% reduction in efficiency.
The band cut and the timing cut each have a slightly different
impact on calibration data and simulated events.
By taking the difference of their acceptance
we evaluated their systematic impact on the signal efficiency
to $\pm 4.2$\% for the band cut and $\pm^{4.2}_{5.1}$\% for the timing cut.
The 1\,$\sigma$ uncertainty in the effective light yield
$\mathcal{L}_{\rm eff}$ as evaluated in \cite{Leff} changes the signal efficiency in a range from $+1.4$\% to $-0.2$\%.
For the decay constants of scintillation light caused by nuclear recoils
we took the uncertainty in the determination of the constants
and the difference between our values and the NEST model \cite{NEST},
$\pm 1$\,ns, as our systematic uncertainty.
The total systematic error on the signal efficiency
for a 50\,GeV WIMP is evaluated to $\pm^{13}_{16}$\%, summing up 
in quadrature the systematic errors as detailed above.
This evaluation was repeated for WIMP masses of
20, 100 and 300\,GeV and 1, 3, and 5\,TeV,
and is assumed to be applicable for masses close to the ones evaluated.

\begin{figure}[]
    \begin{center}
\includegraphics[trim=0cm 0cm 0cm 0cm,width=12cm,height=12cm,angle=0,keepaspectratio,clip]{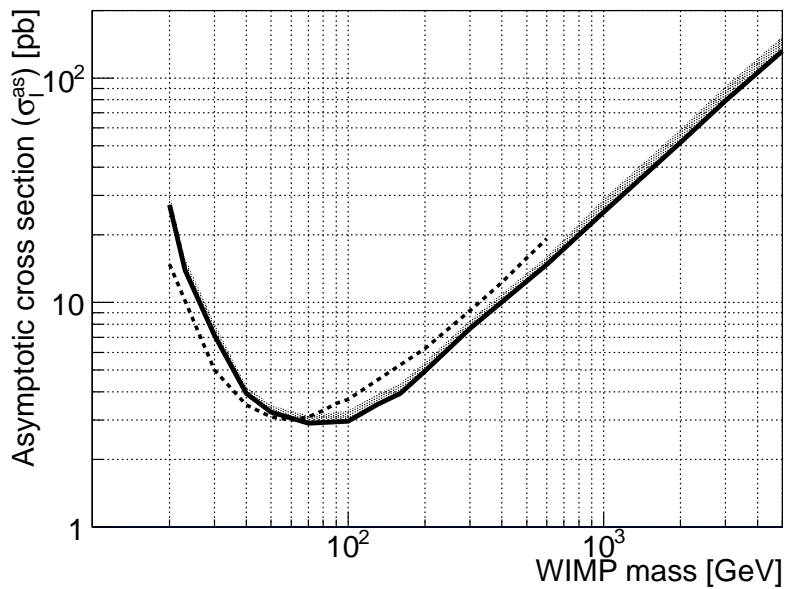}
     \end{center}    
     \caption{The black solid line is our 90\% C.L. upper limit on the asymptotic cross section $\sigma_I^{as}$ for inelastic scattering on $^{129}$Xe (black solid line)
using the same form factors as DAMA.
 The gray band covers its variation with our  systematic uncertainty. The dotted line is the limit obtained by the DAMA group \protect{\cite{dama1996,dama2000}}.
It was derived after statistically subtracting background. Our low background allows us to derive this limit without such background subtraction.
}
     \label{fig:4}
\end{figure}

\begin{figure}[]
    \begin{center}
\includegraphics[trim=0cm 0cm 0cm 0cm,width=10cm,height=10cm,angle=0,keepaspectratio,clip]{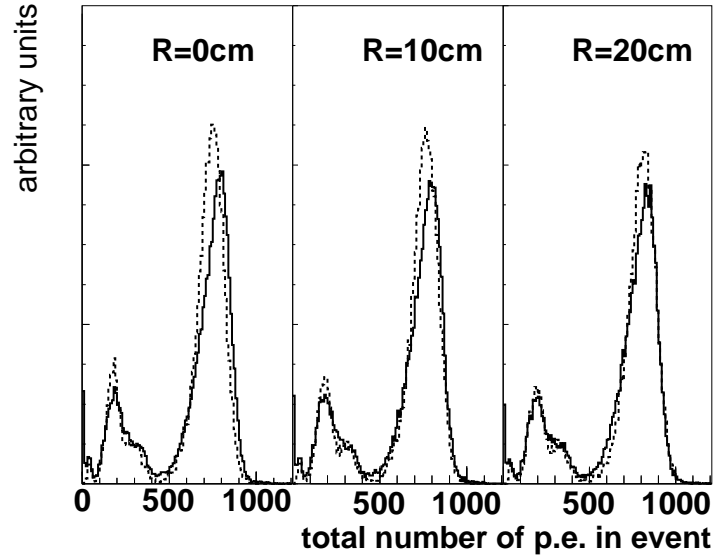}
     \end{center}
     \caption{Comparison between data (solid histograms)
     and simulation (dashed histograms)
     of the energy distribution of $^{241}$Am at three radial positions $R$.
     }
    \label{fig:5}
\end{figure}

\begin{figure}[]
    \begin{center}
\includegraphics[trim=0cm 0cm 0cm 0cm,width=10cm,height=10cm,angle=0,keepaspectratio,clip]{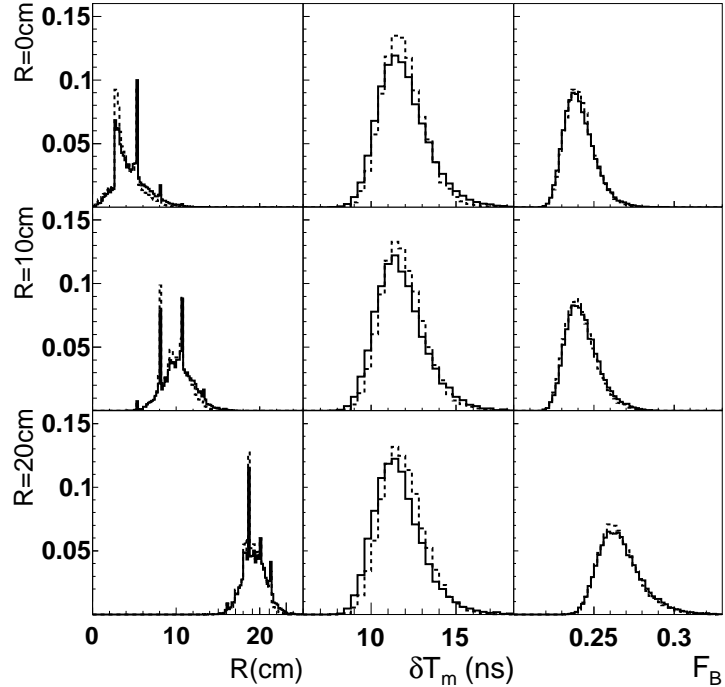}
     \end{center}
     \caption{Comparison between $^{241}$Am data (solid histograms)
     and simulation (dashed histograms)
     for the three cut parameters at three radial positions in the detector,
     $R=0$\,cm, 10\,cm, and 20\,cm, from the top to the bottom, respectively.
     From left to right the distributions for all three parameters, the reconstructed radius, the timing difference, and the band cut parameter, are shown for each of the source positions.}
    \label{fig:6}
\end{figure}

Finally we evaluate
the impact of our systematic uncertainty on the limit
we obtain.
Assuming a true number of events $\mu$ in the energy window before
the event reduction, we can calculate the expected number of observed
events by multiplying with the signal efficiency tabulated in Tab.~\ref{gel1}.
Based on the expected number of observed events, we can generate the
number of observed events following Poisson statistics. This procedure
was repeated to accumulate a histogram of the observed number of events for
a fixed $\mu$ by sampling the signal efficiency within its systematic error.
The 90\% C.L. upper limit for $\mu$ is the one
that results in a 10\% probability to have five events or less.
Using Eq.~6 this is then translated to an inelastic WIMP nucleus
cross section,
and the variation of our limit within our systematic uncertainties.
Both are shown by the black line and gray band in Fig.~\ref{fig:4},
respectively.

It should be noted
that the constraint obtained by the DAMA group \cite{dama1996,dama2000}
was derived from a statistical evaluation of an excess above
a large background of $2\times 10^{-2}$\,keV$^{-1}$d$^{-1}$kg$^{-1}$.
We achieved a lower background
$\sim 3\times 10^{-4}$\,keV$^{-1}$d$^{-1}$kg$^{-1}$
using the cut discussed above.
This low background allowed us to avoid having to subtract background to obtain a competitive limit.

\begin{figure}[]
    \begin{center}
\includegraphics[trim=0cm 0cm 0cm 0cm,width=10cm,height=10cm,angle=0,keepaspectratio,clip]{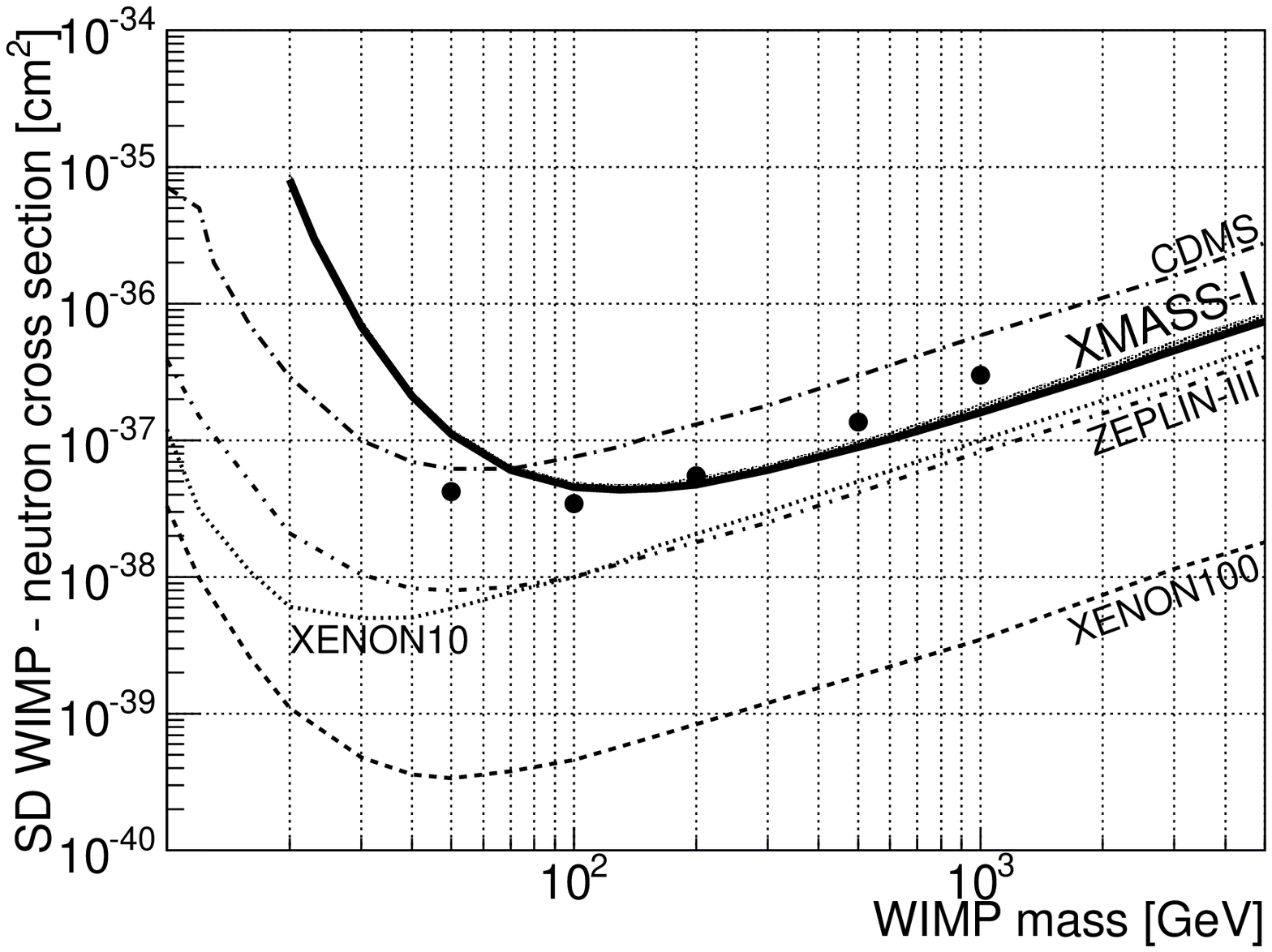}
     \end{center}
     \caption{
	 The thick line with its gray shaded systematic 
	 uncertainty band represents
	 our limit using the form factors of Ref.~\cite{Baudis},
	 and the dots represent our limits following Ref.~\cite{Ejiri2013}
	 for that paper's choice of WIMP masses.
	 The dashed, dotted, dash-dotted, 
	 and long dash-dotted lines represent
	 experimental constraints on spin dependent WIMP nucleon
	 cross sections extracted from elastic scattering data
	 as published in Refs.~\cite{xenon100sd, xenon10sd, zep, CDMSsd}
	 respectively.
	 Our own limit is the first derived exclusively from data
	 on inelastic scattering.
     }
    \label{fig:7}
\end{figure}

Our limits that are based on the updated nuclear form factors
given in Refs.~\cite{Baudis, Ejiri2013} are also using the same
data reduction and therewith event sample as those for the older form
factors. The results shown in Fig.~\ref{fig:7} are based on 
the inelastic structure factors for $S_n(u)$ 1b+2b currents as shown 
in Fig.~1 of Ref.~\cite{Baudis}, and on the time averaged
differential scattering rate for inelastic scattering
as shown in Fig.~4 of Ref.~\cite{Ejiri2013}, the latter taken with
a normalization of 17\,fb\footnote{This value was taken from
Ref.~\cite{Ejiri2013}'s Ref.~[46] and confirmed with one of
the authors of Ref.~\cite{Ejiri2013}.}
for the total WIMP-nucleon
cross section.
Figure \ref{fig:7} also shows previous experimental limits
for the spin-dependent WIMP-neutron cross section
that were previously obtained by XENON10, CDMS, ZEPLIN-III, and XENON100,
Refs.~\cite{xenon100sd, xenon10sd, zep, CDMSsd}.
All of these existing limits are based on spin-dependent
elastic scattering, while our analysis explicitly restricts
itself to inelastic scattering on nucleus.

\begin{table}[]
   \caption{Signal efficiencies with their systematic errors
  for deriving the limit shown in Figs.~\protect{\ref{fig:4}} and \protect{\ref{fig:7}}.
    The row starting from (a) is based on Ref.~\protect{\cite{Engel}},
    and the one starting from (b) on Ref.~\protect{\cite{Baudis}}.
  }
  \label{gel1}
 \begin{center}
  \begin{tabular}{|l|c|c|c|c|c|c|c|}
  \hline
WIMP mass (GeV) & 20 & 50 & 100 & 300 & 1000 & 3000 & 5000\\ \hline
(a) signal efficiency (\%) & $23\pm^7_6$ & 29$\pm^4_5$ & 26$\pm^2_4$ & 19$\pm^1_3$ &16$\pm^1_3$ & 15$\pm^1_3$ & 15$\pm^1_3$\\ \hline
(b) signal efficiency (\%) & $24\pm^7_6$ & 30$\pm^2_5$ & 29$\pm^2_4$ & 26$\pm^2_5$ &25$\pm^2_5$ & 25$\pm^2_5$ & 25$\pm^2_5$\\
  \hline 
  \end{tabular}
  \end{center}
\end{table}

\begin{table}[]
   \caption{Systematic error of the signal efficiency for 
  different WIMP masses.
    As in the previous table (a) is for the signal calculation
    based on Ref.~\cite{Engel} and (b) on Ref.~\cite{Baudis}.
  All the entries are in \%
  of the nominal efficiencies.
  See text for detail.
  $^{241}$Am data and simulated events (see text).}
  \label{gel2}
 \begin{center}
  \begin{tabular}{|l|c|c|c|c|c|c|c|}
  \hline
(a) WIMP mass (GeV) & 20 & 50 & 100 & 300 & 1000 & 3000 & 5000\\ \hline
Energy scale & $\pm^{30}_{22}$ & $\pm^{11}_{13}$ & $\pm^0_{5.1}$ & $\pm^{0.4}_{7.1}$ & $\pm^{1.1}_{9.5}$ & $\pm^{2.2}_{11}$ & $\pm^{2.7}_{11}$\\ \hline
Energy resolution & $\pm^0_{8.2}$  & $\pm^0_{5.6}$ & $\pm^0_{6.8}$ & $\pm^0_{8.1}$ & $\pm^0_{9.7}$ & $\pm^0_{8.8}$ & $\pm^0_{9.0}$\\ \hline
Radius cut & $\pm^0_{3.3}$ & $\pm^0_{3.2}$ & $\pm^0_{4.0}$ & $\pm^0_{5.2}$ & $\pm^0_{6.8}$  & $\pm^0_{6.3}$  & $\pm^0_{6.4}$\\ \hline
Timing cut &$\pm^{4.2}_{5.1}$ &$\pm^{4.2}_{5.1} $ &$\pm^{4.2}_{5.1} $ &$\pm^{4.2}_{5.1} $ &$\pm^{4.2}_{5.1} $ &$\pm^{4.2}_{5.1} $ &$\pm^{4.2}_{5.1} $\\ \hline
Band cut &$\pm 4.2$ &$\pm 4.2$ &$\pm 4.2$ &$\pm 4.2$ &$\pm 4.2$ &$\pm 4.2$ &$\pm 4.2$\\ \hline
$\mathcal{L}_{\rm eff}$ &$\pm^{6.4}_0$ & $\pm^{1.4}_{0.2}$ & $\pm^0_{1.4}$ & $\pm^{3.9}_0$ & $\pm^{1.0}_{1.3}$ & $\pm^0_{1.7}$ & $\pm^0_{4.0}$\\ \hline
$\tau_{nr}$ & $\pm^{0}_{0.8}$ & $\pm^{1.3}_{2.2}$ & $\pm^{0}_{8.8}$ & $\pm^{0}_{4.6}$ & $\pm^{0}_{4.7}$ & $\pm^{0}_{5.7}$ & $\pm^{0}_{5.7}$ \\ \hline
total systematic error & $\pm^{31}_{25}$ & $\pm^{13}_{16}$ & $\pm^{5.9}_{15}$ & $\pm^{7.1}_{14}$ & $\pm^{6.0}_{17}$ & $\pm^{6.3}_{18}$ & $\pm^{6.5}_{18}$\\
\hline 
  \end{tabular}

  \vspace*{3mm}
  \begin{tabular}{|l|c|c|c|c|c|c|c|}
  \hline
(b) WIMP mass (GeV) & 20 & 50 & 100 & 300 & 1000 & 3000 & 5000\\ \hline
Energy scale & $\pm^{27}_{21}$ & $\pm^{5.6}_{9.9}$ & $\pm^{0}_{7.8}$ & $\pm^{4.8}_{13}$ & $\pm^{5.5}_{14}$ & $\pm^{5.9}_{14}$ & $\pm^{6.1}_{17}$\\ \hline
Energy resolution & $\pm^0_{7.1}$  & $\pm^0_{5.7}$ & $\pm^0_{7.9}$ & $\pm^0_{8.9}$ & $\pm^0_{9.7}$ & $\pm^0_{9.5}$ & $\pm^0_{9.5}$\\ \hline
Radius cut & $\pm^0_{3.7}$ & $\pm^0_{4.1}$ & $\pm^0_{4.4}$ & $\pm^0_{4.3}$ & $\pm^0_{5.3}$  & $\pm^0_{4.5}$  & $\pm^0_{3.9}$\\ \hline
Timing cut &$\pm^{4.2}_{5.1}$ &$\pm^{4.2}_{5.1} $ &$\pm^{4.2}_{5.1} $ &$\pm^{4.2}_{5.1} $ &$\pm^{4.2}_{5.1} $ &$\pm^{4.2}_{5.1} $ &$\pm^{4.2}_{5.1} $\\ \hline
Band cut &$\pm 4.2$ &$\pm 4.2$ &$\pm 4.2$ &$\pm 4.2$ &$\pm 4.2$ &$\pm 4.2$ &$\pm 4.2$\\ \hline
$\mathcal{L}_{\rm eff}$ &$\pm^{0}_{2.0}$ & $\pm^{0}_{1.5}$ & $\pm^{1.8}_{0}$ & $\pm^{5.7}_0$ & $\pm^{1.0}_{0}$ & $\pm^{2.2}_{2.9}$ & $\pm^{5.1}_{0}$\\ \hline
$\tau_{nr}$ & $\pm^{0}_{0.8}$ & $\pm^{0}_{4.7}$ & $\pm^{0}_{5.4}$ & $\pm^{0}_{4.7}$ & $\pm^{0.2}_{4.1}$ & $\pm^{0}_{7.1}$ & $\pm^{0}_{4.9}$ \\ \hline
total systematic error & $\pm^{28}_{24}$ & $\pm^{8.2}_{15}$ & $\pm^{6.2}_{15}$ & $\pm^{9.5}_{18}$ & $\pm^{8.2}_{19}$ & $\pm^{8.7}_{20}$ & $\pm^{9.9}_{21}$\\
  \hline 
  \end{tabular}
  \end{center}
\end{table}

\section{Conclusion}
A search for inelastic scattering of WIMPs on ${}^{129}$Xe
was performed using data from our single phase liquid xenon detector XMASS.
Events reconstructed in a spherical fiducial volume of 15\,cm
radius at the center of the detector containing 41\,kg of LXe
were used in this analysis.
We observed no significant excess in 
165.9\,live\,days\rq{} data
and derived for e.g.\ a 50\,GeV WIMP an upper
limit
at the 90\% confidence level
for its inelastic cross section on $^{129}$Xe
nuclei of 3.2\,pb
using the form factors of Ref.~\cite{Engel}
and an upper limit for the spin-dependent WIMP-neutron
cross section of 110\,fb or 42\,fb
respectively using
the updated form factors from Refs.~\cite{Baudis} or \cite{Ejiri2013}.
%

\ack
We gratefully acknowledge the cooperation of Kamioka Mining
and Smelting Company. 
This work was supported by the Japanese Ministry of Education,
Culture, Sports, Science and Technology, Grant-in-Aid
for Scientific Research, and partially
by the National Research Foundation of Korea Grant funded
by the Korean Government (NRF-2011-220-C00006).
We thank Dr.~Masahiro Ibe for useful discussion.

\bibliographystyle{model1-num-names}
\bibliography{<your-bib-database>}


\end{document}